\documentclass[twocolumn,aps,prl,superscriptaddress,showpacs]{revtex4}

\usepackage{graphicx}

\usepackage{amsmath}
\usepackage{epsfig}

\begin{document}
 \title{Experimental demonstration of continuous variable quantum erasing}

\author{Ulrik L. Andersen}
\affiliation{Institut f\"{u}r Optik, Information und Photonik,  Max-Planck Forschungsgruppe Universit\"{a}t Erlangen-N\"{u}rnberg, Staudtstr. 7/B2, 91058, Erlangen, Germany}
\author{Oliver Gl\"ockl}
\affiliation{Institut f\"{u}r Optik, Information und Photonik,  Max-Planck Forschungsgruppe Universit\"{a}t Erlangen-N\"{u}rnberg, Staudtstr. 7/B2, 91058, Erlangen, Germany}
\author{Stefan Lorenz}
\affiliation{Institut f\"{u}r Optik, Information und Photonik,  Max-Planck Forschungsgruppe Universit\"{a}t Erlangen-N\"{u}rnberg, Staudtstr. 7/B2, 91058, Erlangen, Germany}
\author{Gerd Leuchs}
\affiliation{Institut f\"{u}r Optik, Information und Photonik,  Max-Planck Forschungsgruppe Universit\"{a}t Erlangen-N\"{u}rnberg, Staudtstr. 7/B2, 91058, Erlangen, Germany}
\author{Radim Filip}
\affiliation{Department of Optics, Research Center for Optics, Palacky University, 17. Listopadu 50, 77200 Olomouc
Czech Republic}

\date{\today}

\begin{abstract}
We experimentally demonstrate the concept of continuous variable quantum erasing. The amplitude quadrature of the signal state is labelled to another state via a quantum nondemolition interaction, leading to a large uncertainty in the determination of the phase quadrature due to the inextricable complementarity of the two observables. We show that by erasing the amplitude quadrature information we are able to recover the phase quadrature information of the signal state. 
\end{abstract}

\maketitle

Niels Bohr's famous complementarity principle states that simultaneous knowledge of two complementary variables is impossible \cite{bohr28.nat}. The canonical example being the double slit experiment where the determination of which-way knowledge and the observation of interference fringes are mutually exclusive. Any attempt made to measure which way the particles took ultimately destroys the interference pattern.  
However, Scully and Dr\"uhl \cite{scully82.pra} proposed that under certain circumstances, the interference can be fully recovered by erasing the distinguability information. In other words, if one somehow manages to change the measurement strategy such that the which-way information becomes inaccessible, one can trade it for a revival of the interference pattern. 

Experimental verification of complementarity followed by quantum erasing has so far been restricted to the binary quantum variable regime, where only a two-dimensional Hilbert space, corresponding to two different paths (e.g. two paths in an interferometer), has been considered \cite{schwindt99.pra}. However, little attention has been paid to the continuous variable (cv) regime in which the Hilbert space is infinite-dimensional and hence the number of possible "paths" is infinite \cite{filip03.pra}. In this Letter we present the first experimental realization of quantum erasing in a cv setting. Besides being a fundamentally interesting result, the concept of cv quantum erasing may be a useful operation in quantum information processing. The technique may allow quantum states, which have been disturbed during data storage in quantum memories, to be restored \cite{marek04.xxx}. More specifically, a stored quantum state that leaks through an imperfect part of a memory cell can be reconstructed by monitoring certain properties of the leaking state. Furthermore the quantum erasure can be also employed as an in-line squeezing source possesing high coupling efficiencies to an arbitrary input state \cite{radim04}.

 


The complementary pair with a continuous spectrum that we will consider is the pair of canonically conjugate quadrature amplitudes: the amplitude and phase quadratures of light.
The complementarity arises from the intrinsic impossibility of gaining perfect information about the two conjugate quadratures as formulated in the generalized Heisenberg uncertainty relation \cite{arthurs88.prl}. 
Contrary to the binary case, where only two different trajectories are possible (corresponding to two different eigenstates), in the cv case the amplitude quadrature can take on a continuous set of possible eigenvalues. In an act of discriminating between these eigenstates ("paths"), referred to as {\it Which Eigenstate} (WE) information \cite{bjork99b.pra}, the probability distribution of the complimentary variable, here the phase quadrature, is broadened. This is the analog to "washing out" the interference pattern in the binary setting.   

A way of gaining WE information about the amplitude quadrature is to encode the signal information into another state, called the marker state, using a quantum nondemolition (QND) entangling coupling\cite{braginsky}. Such a coupling has several times been demonstrated with special reference to the execution of QND measurements\cite{braginsky,grangier98.nat}.
Due to the mere possibility of acquiring precise knowledge about the amplitude quadrature of the signal state, an enlarged uncertainty in the determination of the conjugate quadrature variable is inevitable.
In this Letter we show that it is however possible to reverse this process and revive the conjugate information. The price one has to pay is an {\it erasing} of the WE information that was extracted using the QND interaction. This is the principle of quantum erasure. 



We consider two states, the signal state, $s$ and the marker state, $m$, described in the infinite-dimensional Hilbert space ${\cal H}={\cal H}_s\otimes{\cal H}_m$ and with corresponding canonical operators, the amplitude quadrature $\hat{x}_{s,m}$ and the phase quadrature $\hat{p}_{s,m}$.
The marker state is described by the coordinate eigenstate $|x=0\rangle _m$ whereas the signal state is described by the coherent state  $|\alpha\rangle _s$. The beams employed in our experiment accommodate Gaussian statistics and it is instructive to write the signal state in the configuration space as $|\alpha\rangle\propto\int{dx f(x)|x\rangle}$ where $|x\rangle$ are the possible eigenstates ("paths") of the signal and $f(x)$ is a state dependent function.
Therefore in the cv regime the different "paths" correspond to internal (not physically separable) orthogonal eigenstates ($\langle x_i| x_j\rangle=\delta_{ij}$) of the signal. 
As in the binary case, the aim is now to discriminate between orthogonal eigenstates of the signal state. This labelling procedure is carried out by a unitary QND interaction (with the unitary operator $U_Q$) acting on the input joint state vector, leaving the output in an entangled state of the form $U_Q|\alpha\rangle_s|x=0\rangle_m$.
Measuring the marker output beam, the signal state is projected onto a certain eigenstate corresponding to the result of the measurement and consequently leading to the destruction of the complementary information. 
However, measuring the complementary variable of the marker output the signal state can be deterministically projected back to the initial state.


In our scheme we use a beam splitter coupled to squeezed light in order to accomplish the QND coupling\cite{bruckmeier97.prl}. The performance of such a QND system is limited by the degree of squeezing which in a realistic experimental situation is never arbitrarily high.
We therefore define a gain normalized variance of the uncertainty associated with the labelling of the signal information onto the marker beam \cite{grangier98.nat}:
$N_{x_s,in\rightarrow x_m,out}=\langle\Delta x_{m,out}^2\rangle/g_{m}^2-\langle\Delta x_{s,in}^2\rangle$
where $\langle\Delta x^2\rangle=\langle x^2\rangle-\langle x\rangle^2$. $\hat{x}_{s,in}$ and $\hat{x}_{m,out}$ are the amplitude quadrature operators of the signal input and marker output respectively and $g_m$ is the gain associated with the signal to marker transformation. For our approach this gain is directly related to the transmission coefficient, $T$, of the QND beam splitter, $g_m=\sqrt{1-T}$.
Similarly, the gain normalized added uncertainty in the conjugate quadrature of the signal is
$N_{p_s,in\rightarrow p_s,out}=\langle\Delta p_{s,out}^2\rangle/g_{s}^2-\langle\Delta p_{s,in}^2\rangle$
where $g_s=\sqrt{T}$ is the signal gain.
Thus knowing the gain of the apparatus together with the input and output noise levels, the induced measurement uncertainty can be quantified in terms of the added broadening of the signal uncertainty distribution. We derive the uncertainty product \cite{grangier98.nat}
$N_{x_s,in\rightarrow x_m,out}N_{p_s,in\rightarrow p_s,out}\geq 1$
which manifests the complementarity between the two quadrature variables. 

By allowing for a local unitary operation on the marker output state so that the amplitude quadrature information is completely inaccessible to our detection system, the WE information is erased. Principally this is done by measuring the phase quadrature which yields no information about the amplitude quadrature. Using such a measurement strategy we may conditionally recover the phase quadrature information of the signal state by displacing the signal output operator $\hat{p}_{s,out}$ according to the measurement result of the marker output $p_{m,out}$ scaled by an appropriate gain factor $G$. The conditioned result can be formulated as $\hat{p}_{c}=\hat{p}_{s,out}-Gp_{m,out}$.
The efficiency of the state restoration can again be quantified in terms of the induced broadening of the distribution function with respect to the original signal state:
$N_{p_s,in\rightarrow p_c}=\langle\Delta p_{c}^2\rangle/g_{e}^2-\langle\Delta p_{s,in}^2\rangle$
which is also normalized to the overall gain, $g_e=1/\sqrt{T}$, of the process, now including the conditioned measurement stage. A similar expression holds for the conditioned amplitude quadrature but with a normalization gain of $1/g_e$.
These gains are included to ensure complete quantum state restoration; physically they correspond to unitary local squeezing operations.



\begin{figure}[h] \centering \includegraphics[width=6cm]{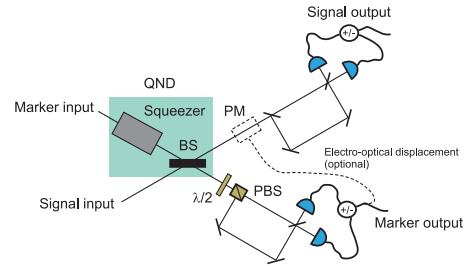} \caption{\it Schematic of the cv quantum erasing scheme. BS: Beam splitter, PBS: Polarizing beam splitter and PM: Phase modulator.}
\label{fig1}  \end{figure}

A schematic diagram of our setup is shown in Fig.~\ref{fig1}. The center piece is the QND device: a beam splitter where one input port is illuminated by the squeezed marker beam \cite{bruckmeier97.prl}. 
Femtosecond light pulses at 1530 nm, generated by downconverting the output from a mode-locked Ti:sapphire laser, are injected into a fiber Sagnac interferometer to generate amplitude squeezed light pulses \cite{schmitt98.prl}. The signal beam under investigation is combined with these squeezed light pulses at the beam splitter, creating signal and marker output beams. 
The illustration of complementarity and erasure is restricted to a certain spectral mode (sideband) with the frequency $\Omega =20.5 MHz$. The squeezing spectrum consists of a very bright component at the laser carrier frequency, which acts as an internal local oscillator to probe the sidebands. 

To quantify the added noise contributions we must be able to measure the amplitude as well as the phase quadratures of the two output beams. Amplitude quadrature information is easily acquired by direct detection. Phase quadrature measurements of bright beams are more involved. To accomplish such a measurement we employed an interferometric setup capable of 
rotating our intrinsic local oscillator with respect to the quantum noise sidebands such that the phase quadrature is mapped onto the amplitude quadrature and is consequently measurable \cite{glockl}. 
This is done by injecting the beam into a Mach-Zehnder interferometer with strongly unbalanced arm lengths (see Fig.~\ref{fig1}). 
Two balanced detectors measure the output beams to form the photocurrent difference. The overall efficiency of the detection system including detector and mode matching efficiencies is 70\%.
For our measurement frequency, accurate mapping of the conjugate quadratures is obtained by choosing an arm length difference of 7.3 m corresponding to a phase shift of $\pi$ at the measurement frequency, while the optical phase shift was $\pi/2$. The first beam splitter in the marker beam interferometer consists of a half-wave plate in combination with a polarizing beam splitter. This allow switching between a phase quadrature measurement and an amplitude quadrature measurement by setting the wave plate angle to $22.5^{\circ}$ and $0^{\circ}$, respectively. A simple unitary operation therefore enables conjugate quadratures of a bright beam to be measured.    

\begin{figure}[h] \centering \includegraphics[width=6.5cm]{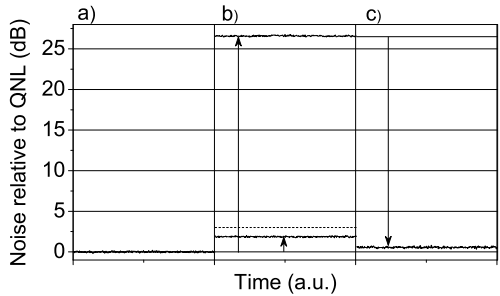} \caption{\it Experimental results for the cv quantum erasing concept. a) Shows the quantum noise level of the input state, b) is the phase quadrature noise level of the signal output (upper trace) and the amplitude quadrature noise of the marker output (lower trace) and c) is the conditioned phase quadrature noise level. Note that the noise levels have been normalized by the respective gains. The dashed line represents the noise level expected for a classical device corresponding to no squeezing. Resolution band width: 300kHz Video band width: 30Hz. Sweep time: 5 sec.}
\label{fig2}  \end{figure}

The experimental verification of cv quantum erasing consists of four steps: 1) Preparation of an input state, 2) the QND interaction, 3) erasing and 4) revival of the phase information. 1) The input signal state was chosen to be the vacuum state to avoid complications in the detection process. By this choice we are confident that it is a pure state displaying an uncertainty at the standard quantum level ($\langle\Delta x_{in}^2\rangle=\langle\Delta p_{in}^2\rangle=1$). In column a of Fig. \ref{fig2} the noise power of this input state is shown. We note that there is no need of setting up the signal state to generate certain eigenstates (like in the double slit experiment) since such different eigenstates are automatically incorporated in the cv Gaussian signal state. 2) The input signal is now superposed with a bright squeezed beam on a beam splitter hereby labelling the amplitude quadrature information of the signal state onto the marker output state. 
Ideally, the noise level of the marker output (scaled to the gain of the beam splitter) is identical to the noise level of the signal input meaning that the eigenstates of the signal state are perfectly tracked. But due to the imperfect squeezing there is a small error in the determination of the amplitude quadrature and consequently a raise in the marker noise level with respect to the signal input noise level is observed. This is measured by directly detecting the amplitude of the marker output beam, analyzing the photocurrents in a spectrum analyzer and normalizing the result by the gain $g_m=\sqrt{1-T}$, where $T=0.477$ is the transmittivity of the QND beam splitter. The result is shown by the lower trace in column b of Fig.~\ref{fig2}. We find the added noise to be $N_{x_s,in\rightarrow x_m,out}=0.55\pm0.02$, which is below the unity value that one would expect in a classical scenario. 
As shown by the upper trace in column b the phase quadrature is now correspondingly increased with an added noise of $N_{p_s,in\rightarrow p_s,out}=455\pm7$ (normalised to $g_s=\sqrt{T}$). The broadening of the phase distribution is not solely a result of the labelling procedure but also a result of classical noise invasion introduced in the optical fiber \cite{shelby86.prl}. 3) We now perform a phase quadrature measurement of the output marker beam, i.e. in a base where the WE information is completely erased and the maximally incompatible information becomes accessible. 4) Joint measurements of the phase quadratures now allows the phase information of the original signal state to be revealed. Technically this is done by subtracting the photocurrents produced by the phase quadrature measurement of the signal output with those of the marker output. The noise level of this conditioned output state normalised to the gain $g_e=\sqrt{1/T}$ is shown in column c of Fig. \ref{fig2}. We clearly see that the erasing procedure almost recovers the original noise level of the signal input, the resulting noise level being only $0.56\pm0.08dB$ ($N_{p_s,in\rightarrow p_c}=0.14\pm0.02$) above the quantum noise level.  The slight discrepancy from the optimum value of 0dB can be explained by slight inefficiencies in the balancing process between the four photo detectors. We stress that the first two steps are executed in order to verify the validity of the QND coupling and basically constitute a normal QND measurement, whereas the last two steps demonstrate the essence of the experiment namely a demostration of cv quantum erasing.

\begin{figure}[h] \centering \includegraphics[width=6.5cm]{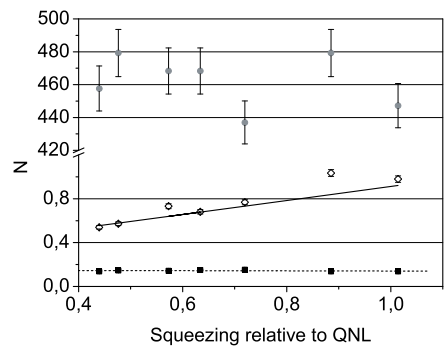} \caption{\it Added noise values for the amplitude quadrature of the marker output without erasing $N_{x_s,in\rightarrow x_m,out}$ (open diamonds), the phase quadrature of the signal output without erasing $N_{p_s,in\rightarrow p_s,out}$ (circles), and the phase quadrature of the signal output with erasing $N_{p_s,in\rightarrow p_c}$ (squares). The dashed line is best linear fit to the measured values.}
\label{fig3}  \end{figure}

Fig. \ref{fig3} shows the results for the noise powers with and without the erasing process associated with seven different squeezing values corresponding to different efficiencies of the QND device. 
The measured values of $N_{x_s,in\rightarrow x_m,out}$ (open diamonds) increase when the degree of input squeezing is decreasing whereas $N_{p_s,in\rightarrow p_s,out}$ (circles) varies around a constant value. 
In Fig. \ref{fig3} we also display the set of data obtained for the joint phase quadrature measurements. These data show, as expected, that the recovery of phase information is independent on the efficiency of the QND device. 
The solid line in Fig. \ref{fig3} is an estimation of the added noise variance based upon the measured propagation losses and detector efficiencies.

The system contains the most important aspect of a quantum erasing experiment, namely an element of delayed choice.
Since the amplitude information is measured by a distinct detector system, the decision of whether to measure the amplitude information or to erase it and instead measure the phase information can be performed {\it after} the detection of the signal beam. The marker state can be stored in a memory cell and the experimenter can decide at any instance what he wants to acquire: knowledge about the amplitude or the phase.

\begin{figure}[h] \centering \includegraphics[width=8.5cm]{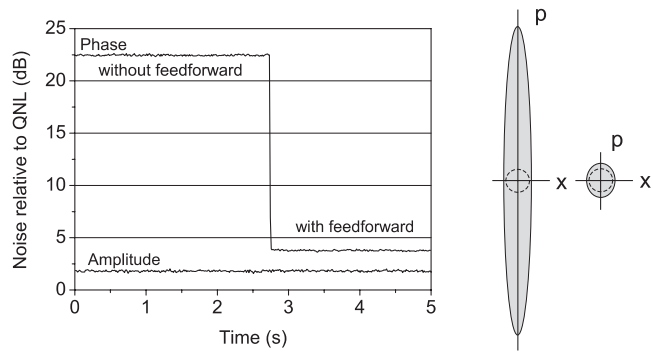} \caption{\it Experimental results for quantum erasing using an electro-optic feed forward loop. a) Shows the noise traces of the signal output before and after the erasing operation and b) shows the estimated contours of the Wigner funtions. Resolution band width: 300kHz Video band width: 30Hz.}
\label{fig4}  \end{figure}

In analogy to previous erasure experiments in the discrete variable regime, the displacement operation in the above mentioned experiment was performed purely electronically. However, in terms of practical applications it would be interesting to perform the displacement with an electro-optic feed forward loop whereby the initial signal is restored as a freely propagating state. An implementation of this idea is shown by the dashed lines in Fig.~\ref{fig1}. Here the correlation procedure (step 4)) is performed by displacing the signal state (via a phase modulator PM) according to the measurement outcome obtained by detecting the marker state after erasing.
The results of such an experiment is presented in Fig.~\ref{fig4} which shows the noise traces of the amplitude and phase quadratures of the output signal state for different operating conditions together with the inferred contours for the Wigner functions. The variances of the signal output beam have been appropriately renormalized according to the gain of the beam splitter. The evolution of the signal state is as follow: before the QND interaction the state is quantum noise limited for all quadratures, after the QND action the complementarity is clearly displayed by the huge phase quadrature uncertainty and finally the QND interaction is reversed using the feedforward loop hereby producing a faithful copy of the input state, only limited by the imperfect squeezing of the marker beam and technical imperfections of the feed forward loop. The added noise variances, with the same $g_e$ as above, were measured to be $0.54\pm0.02$ and $1.39\pm0.03$ for the amplitude and phase quadratures, respectively, and the fidelity between the input signal and the restored output signal was found to be $F=0.68\pm0.01$.


In the presented experiment, we used light beams to perform Which Eigenstate and quantum erasing experiments in the continuous variable regime. Using two mutually exclusive measurement bases we swapped between attaining good WE information and almost perfect information about the conjugate variable. We also demonstrated reconstruction of a state that was detoriated by a QND interaction. Such techniques are particularly useful for protecting quantum states in quantum memories to combat decoherence.

This work has been supported by the Schwerpunkt programm 1078 of the Deutsche Forschungsgemeinschaft, the network of competence QIP of the state of Bavaria (A8), grant 202/03/D239 of Czech Grant Agency and project
LN00A015 and CEZ:J14/98 of the Ministry of Education of Czech Republic. We thank J. Heersink for help with the manuscript. ULA also acknowledges funding from the Alexander von Humboldt Foundation. 









\begin{thebibliography}{}

\bibitem{bohr28.nat} N. Bohr, Nature {\bf 121}, 580 (1928) 
\bibitem{scully82.pra} M.O. Scully and K. Dr\"uhl, Phys. Rev. A {\bf 25}, 2208 (1982).  M.O. Scully et al., Nature (London) {\bf 351}, 111 (1991)
\bibitem{schwindt99.pra} T.J. Herzog et al., Phys. Rev. Lett. {\bf 75}, 3034 (1995). P.D.D. Schwindt et al., Phys. Rev. A {\bf 60}, 4285 (1999). P. Bertet et al., Nature {\bf 411}, 166 (2001). Y.H. Kim et al., Phys. Rev. Lett. {\bf 84}, 1 (2000). T. Tsegaye et al., Phys. Rev. A {\bf 662}, 032106 (2000). S. D\"urr et al, Nature (London) {\bf 395}, 33 (1998). A. Trifonov et al., Eur. Phys. J. D {\bf 18}, 251 (2002). T.B. Pittman et al., Phys. Rev. Lett. {\bf 77}, 1917 (1996). X.Y. Zou et al., Phys. Rev. Lett. {\bf 67}, 318 (1991). A.G. Zajonc et al., Nature (London) {\bf 353}, 507 (1991).
\bibitem{filip03.pra} R. Filip, Phys. Rev. A {\bf 67}, 042111 (2003)
\bibitem{marek04.xxx} P. Marek and R. Filip, Phys. Rev. A, {\bf 70} 022305 (2004). 
\bibitem{radim04} R. Filip and P. Marek, In preparation (2004).
\bibitem{arthurs88.prl} E. Arthurs and M.S. Goodman, Phys. Rev. Lett. {\bf 60}, 2447 (1988)
\bibitem{bjork99b.pra} G. Bj\"ork et al., Phys. Rev. A {\bf 60}, 1874 (1999). G. Bj\"ork and A. Karlsson, Phys. Rev. A {\bf 58}, 3477 (1998)
\bibitem{braginsky} V.B. Braginsky et al., Science 209, 547 (1980).
\bibitem{grangier98.nat} Ph. Grangier et al., Nature {\bf 396}, 537 (1998). J.-Ph. Poizat et al., Ann. Phys. Fr. {\bf 19}, 265 (1994).
\bibitem{bruckmeier97.prl} R. Bruckmeier et al., Phys. Rev. Lett. {\bf 79}, 43 (1997)
\bibitem{schmitt98.prl} S. Schmitt et al., Phys. Rev. Lett. {\bf 81}, 2446, (1998)
\bibitem{glockl} O. Gl\"ockl et al.,  Opt. Lett. {\bf 29}, 1936 (2004). 
\bibitem{shelby86.prl} R.M. Shelby, M.D. Levenson and P.W. Bayer, Phys. Rev. Lett. {\bf 54}, 939 (1985)



\end{thebibliography}
\end{document}